\documentclass[twocolumn]{aastex631}

\begin{document}

\title{Searching for radio pulsars in old open clusters from the Parkes archive}

\author{S.B. Zhang}
\affiliation{Purple Mountain Observatory, Chinese Academy of Sciences, Nanjing 210023, China}
\affiliation{CSIRO Space and Astronomy, Australia Telescope National Facility, PO Box 76, Epping, NSW 1710, Australia}

\author{J. J. Wei}
\affiliation{Purple Mountain Observatory, Chinese Academy of Sciences, Nanjing 210023, China}

\author{X. Yang}
\affiliation{Purple Mountain Observatory, Chinese Academy of Sciences, Nanjing 210023, China}
\affiliation{School of Astronomy and Space Sciences, University of Science and Technology of China, Hefei 230026, China}

\author{S. Dai}
\affiliation{CSIRO Space and Astronomy, Australia Telescope National Facility, PO Box 76, Epping, NSW 1710, Australia}

\author{J. S. Wang}
\affiliation{Max-Planck-Institut f\"ur Kernphysik, Saupfercheckweg 1, D-69117 Heidelberg, Germany}

\author{L. Toomey}
\affiliation{CSIRO Space and Astronomy, Australia Telescope National Facility, PO Box 76, Epping, NSW 1710, Australia}

\author{S.Q. Wang}
\affiliation{Xinjiang Astronomical Observatory, Chinese Academy of Sciences, Urumqi, Xinjiang 830011, People's Republic of China}
\affiliation{CSIRO Astronomy and Space Science, PO Box 76, Epping, NSW 1710, Australia}
\affiliation{Key Laboratory of Radio Astronomy, Chinese Academy of Sciences, Urumqi, Xinjiang, 830011, People's Republic of China}

\author{G. Hobbs}
\affiliation{CSIRO Space and Astronomy, Australia Telescope National Facility, PO Box 76, Epping, NSW 1710, Australia}

\author{X. F. Wu}
\affiliation{Purple Mountain Observatory, Chinese Academy of Sciences, Nanjing 210023, China}
\affiliation{School of Astronomy and Space Sciences, University of Science and Technology of China, Hefei 230026, China}

\author{L. Staveley-Smith}
\affiliation{International Centre for Radio Astronomy Research, University of Western Australia, Crawley, WA 6009, Australia}
\affiliation{ARC Centre of Excellence for All Sky Astrophysics in 3 Dimensions (ASTRO 3D)}



\begin{abstract} 
Motivated by the discovery of a pulsar in the direction of the old open cluster NGC 6791, we conducted a search for radio pulsars in archival Parkes observations targeting similar old open clusters. 
We reprocessed 224 observations totalling 75.02 hours from four clusters: Theia 1661, NGC 6259, Pismis 3, and Trumpler 20. Our analysis identified five known pulsars and three new rotating radio transient (RRAT) candidates. By comparing the measured dispersion measures (DMs) with the expected DM values for each cluster derived from YMW16 and NE2001 models, we conclude that most detected sources are likely background pulsars.
However, RRAT J1749$-$25 in Theia 1661 and RRAT J1237$-$60 in Trumpler 20 have DMs reasonably close to their respective clusters, suggesting possible membership. The association between PSR J1750$-$2536 and Theia 1661 remains ambiguous due to its intermediate DM.
These candidate cluster-associated neutron stars warrant follow-up with more sensitive telescopes such as MeerKAT or the SKA, potentially offering valuable insights into neutron star retention mechanisms and evolution in open cluster environments.
\end{abstract}



\section{Introduction} \label{sec:intro}

Open clusters serve as important laboratories for studying stellar evolution, dynamical interactions, and the late stages of stellar life cycles~\citep{van20}. In contrast to globular clusters, which are ancient and dense stellar systems with deep gravitational potential wells, open clusters exhibit lower stellar densities, younger ages, and shallower potential wells~\citep{Liu25}. These environmental differences create significant challenges for neutron star retention and subsequent pulsar detection in open clusters~\citep{Davies98,Fragione20}. 
A key factor contributing to this challenge is that the low escape velocities of open clusters are generally insufficient to retain neutron stars receiving natal kicks during core-collapse supernovae~\citep{Davies98, Pfahl02, Fragione20}. Recent N-body simulations suggest that only neutron stars formed in massive binaries within open clusters have non-negligible retention probabilities, as binary interactions can reduce kick velocities~\citep{Fragione20,Tanikawa24}.

While globular clusters efficiently retain neutron stars~\citep{Davies98} and host over 300 confirmed radio pulsars~\footnote{\url{https://www3.mpifr-bonn.mpg.de/staff/pfreire/GCpsr.html}}, the situation in open clusters is markedly different. To date, only an X-ray pulsar, CXO~J164710.2$-$455216, has been discovered in the young, massive Galactic cluster Westerlund 1~\citep{Muno06}, and one radio pulsar, J1932+1059, has been proposed as a runaway system originating from the open cluster IC~4665~\citep{Bobylev08}. No radio pulsar had been definitively confirmed within any open clusters. 
This striking disparity persists despite comparable initial neutron star formation rates in both environments~\citep{Fragione20,Tanikawa24}, suggesting fundamental differences in their evolutionary pathways. 
However, the recent discovery of a pulsar in the direction of the old open cluster NGC 6791~\citep{Liu25} has renewed interest in searching for radio pulsars in similar environments. Any confirmed pulsar discovery in such clusters would offer crucial insights into their supernova history, neutron star retention mechanisms, and overall dynamical evolution~\citep{Spina22}.

The Parkes 64 m-diameter radio telescope (``Murriyang'') has discovered the largest number of known pulsars through surveys such as the Parkes multibeam pulsar survey~\citep[PMPS;][]{Manchester01, Hobbs04}. 
The majority of the high-time resolution data collected by Parkes is publicly available through CSIRO’s data archive\footnote{CSIRO Data Access Portal, DAP, \url{https://data.csiro.au}}~\citep{Hobbs11}.
This archive provides an extensive dataset spanning over 30 years, recorded using stable observing systems, and serves as a valuable resource for testing new data processing algorithms on large volumes of pulsar observations. 
Recent reprocessing of archival Parkes data using single-pulse search techniques has led to detecting new fast radio bursts (FRBs) and rotating radio transients (RRATs)~\citep{Zhang19,Zhang20,Yang21}.
Similarly, periodicity searches have uncovered 37 new pulsars in the Parkes Multibeam Pulsar Survey (PMPS)~\citep{Sengar23} and 71 additional pulsars in the High Time Resolution Universe (HTRU) pulsar survey of the southern Galactic plane~\citep{Sengar25}.  

In this study, we reprocessed archival search-mode observations~\footnote{The observations were not originally conducted to monitor these specific open clusters; rather, they were carried out under several different project IDs in the past, such as P050, P268, P366, P512, P574, P595, P630, P860 and P1178.} from the Parkes radio telescope to search for pulsars in the direction of old open clusters similar to NGC 6791. 
Our search strategy aimed to identify both periodic signals and single pulses, which could indicate the presence of either regular pulsars or RRATs~\citep{McLaughlin06}.
In this paper, we describe the details of the observations and data reduction in Section~\ref{sec:obs}. The results from these data sets are presented and discussed in Section~\ref{sec:results}. We conclude in Section~\ref{sec:con}.

\section{Observation and data reduction} \label{sec:obs}

\begin{figure*}[!htb]
    \centering
    \begin{tabular}{ll}
    \includegraphics[width=0.48\linewidth]{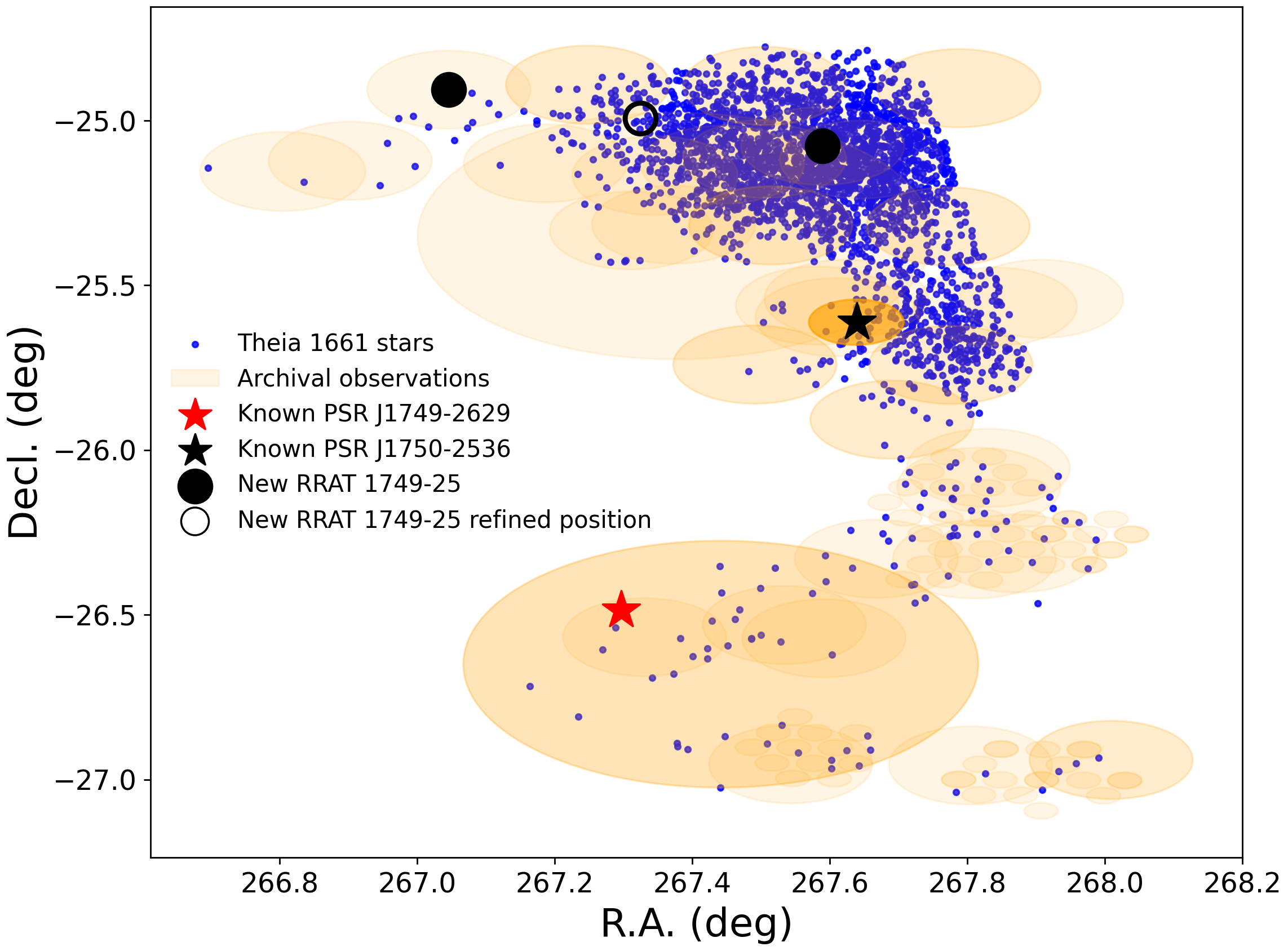} &\includegraphics[width=0.48\linewidth]{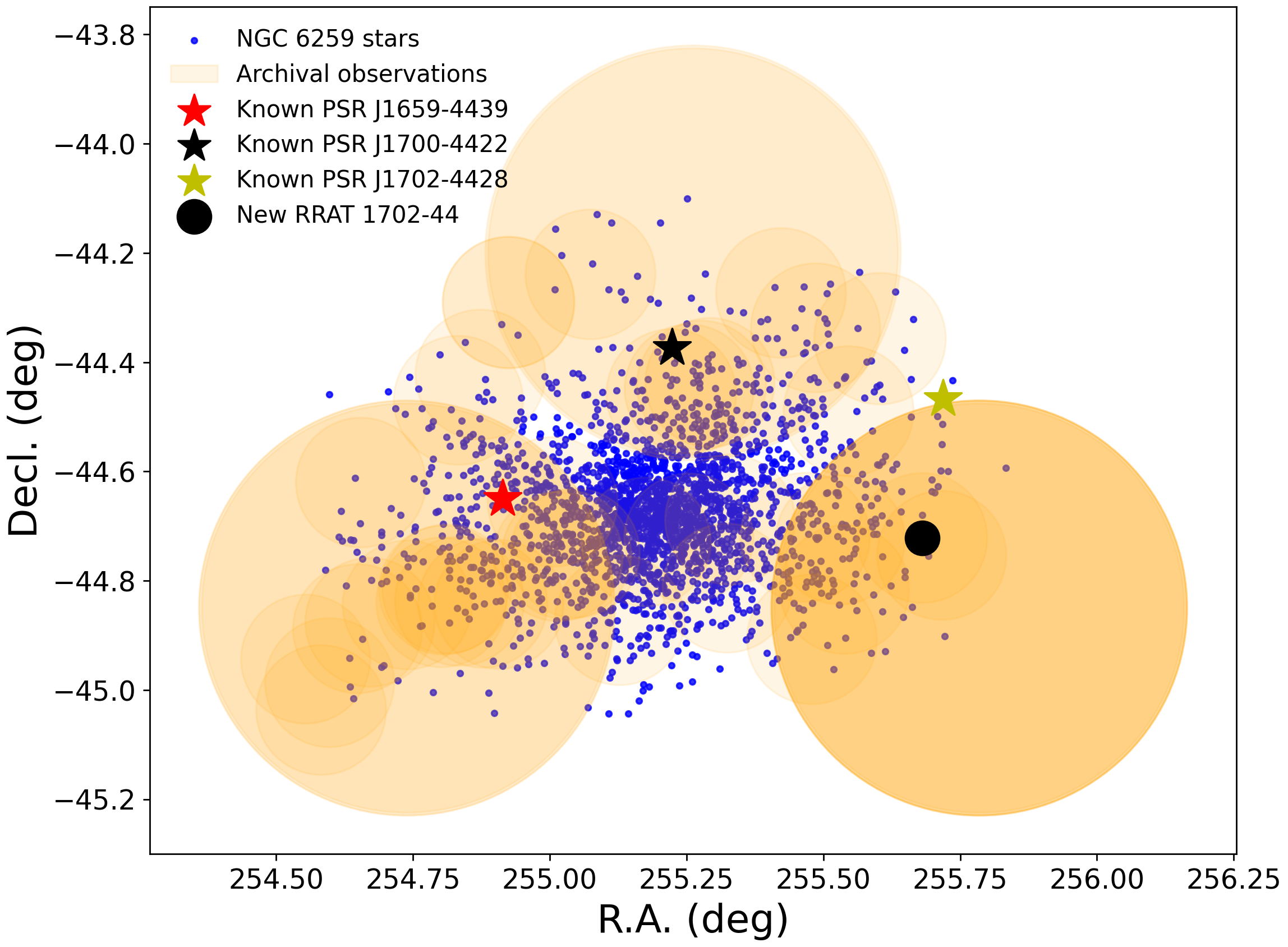}\\
    \includegraphics[width=0.47\linewidth]{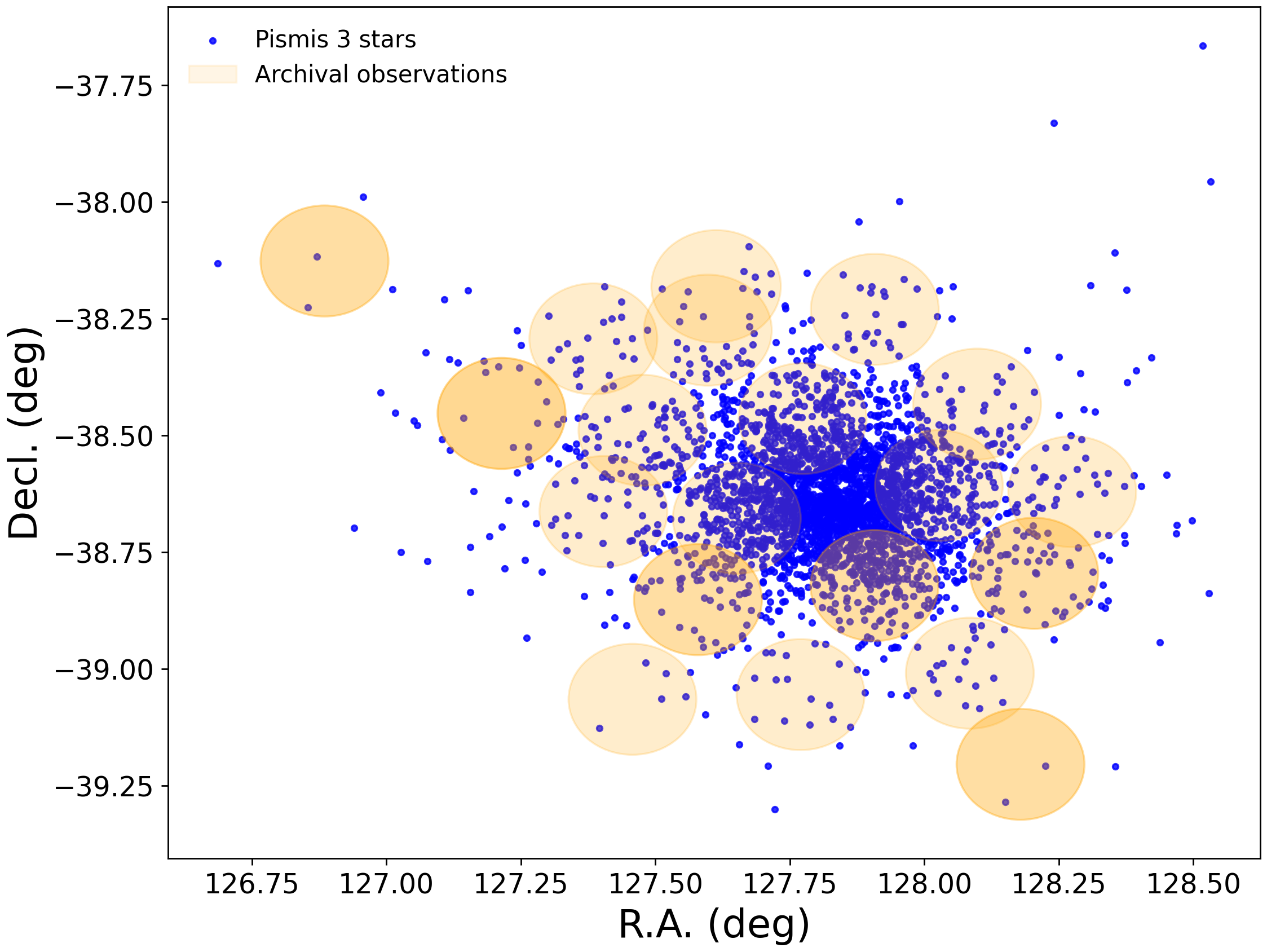} &\includegraphics[width=0.47\linewidth]{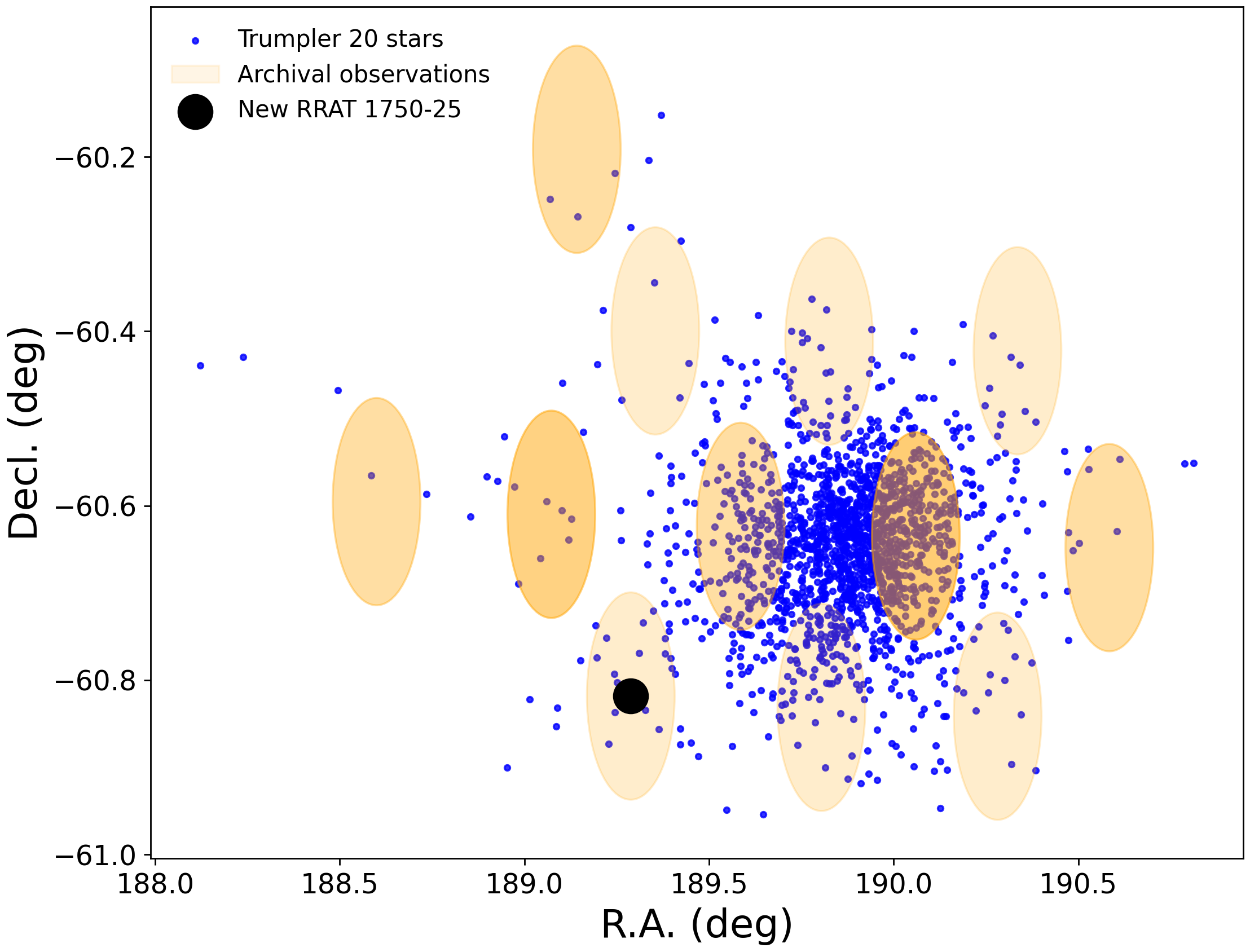}\\
    \end{tabular}
    \caption{Spatial distribution of Parkes archival observations (orange filled circles, diameters indicate the HPBW) overlaid with member stars (blue dots) of four old open clusters: Theia 1661, NGC 6259, Pismis 3, and Trumpler 20.
    All observations are represented as circles with varying radii, but appear elliptical in some panels due to different scaling ratios between the Decl. (y-axis) and R.A. (x-axis) coordinates. Darker orange colouring indicates multiple overlapping observations at the same sky position, with colour intensity proportional to the number of observations.
    Known pulsars are marked with filled stars, while newly discovered RRAT candidates are shown with black filled circles. The refined position for RRAT J1749$-$25 is indicated by an unfilled circle.}
    \label{fig:pointing}
\end{figure*}

To select old open clusters similar to NGC 6791, we followed the selection criteria of~\citet{Liu25}: a total mass exceeding $10^4$~M$_\odot$, a core size smaller than $10$~pc, and an estimated age of at least 100~Myr$-$comparable to the typical lifetimes of canonical pulsars.
Five clusters meet these criteria and are located within the sky coverage of the Parkes radio telescope: Theia 1661, NGC 6259, Pismis 3, Trumpler 20 and Trumpler 5. 
However, no archival Parkes data are available for Trumpler 5.
For the remaining four clusters, a total of 224 archival search-mode observations of 75.02 hours were identified covering the sky positions of their member stars~\citep{Hunt24}, as presented in Figure~\ref{fig:pointing}.
Most of the data were collected using the Parkes 21-cm Multibeam receiver~\citep{Staveley-Smith96MB}, while a few observations were obtained using the 70-cm receiver, the 10/50 cm dual-band receiver, the Methanol Multibeam receiver~\citep{Green12}, or the Ultra-Wideband Low (UWL) receiver~\citep{Hobbs20UWL} system. Among the four clusters, Theia 1661 had the largest archival data sets, with 127 observations of 47.73 hours. NGC 6259, Pismis 3, and Trumpler 20 have 51 observations of 10.05 hours,  26 observations of 5.57 hours, and 20 observations of 11.67 hours, respectively.
Basic properties of these clusters, including their estimated dispersion measures (DMs) from the YMW16~\citep{Yao17} and NE2001~\citep{ne2001} electron density models of the clusters, as well as the related Parkes archival observations, are summarised in Table~\ref{tab:oc_list}.

\begin{table*}
\footnotesize
    \centering
\begin{tabular}{ccrccc|ccccccc}
\toprule
Name  &    R.A.     &  Decl.      &   $d$   &  DM$_{\rm YMW16}$ &  DM$_{\rm NE2001}$  &  Observations          &  Observations &  Known  & New & Detected  \\ 
    &   (deg)   &  (deg)    & (kpc) &  (pc\,cm$^{-3}$) & (pc\,cm$^{-3}$)  & number    & length (hr)    &  sources & sources & sources  \\
    \hline   
Theia 1661  &  267.59   & $-25.15$  &  2.30  &  82.2  & 113.0  & 127  & 47.73 &  2 & 1 & 2 \\      
NGC 6259  &  255.19   & $-44.68$  &  2.16    &  76.6  &  71.6 & 51  & 10.05 &   3 & 1 & 4 \\     
Pismis 3  &  127.83   & $-38.66$  &  2.11    &  157.1  & 173.6  & 26  & 5.57 &  0 & 0&  0   \\     
Trumpler 20  &  189.89   & $-60.63$  &  3.28 &  194.9  & 166.8  & 20  & 11.67  &  0 & 1 & 1 &   \\     
\hline
Total      &             &           &         & &      &       224  &  75.02 &  5 & 3 & 7   \\
\hline
\end{tabular}
    \caption{Properties of selected old open clusters and summary of Parkes archival observations. The table includes cluster coordinates, distances, estimated DMs from YMW16 and NE2001 electron density models, number of observations, total observation time, and detection statistics of known and new sources for each cluster.}
    \label{tab:oc_list}
\end{table*}

Data were processed using search pipelines based on the pulsar/FRB single pulse searching package \emph{\sc presto}~\citep{Ransom01}. 
We divided the UWL data into a series of sub-bands ranging from 128 to 3328\,MHz based on a tiered strategy~\citep{Kumar21_11a}, but processed the full-band data from the observations of the remaining receivers following a well-used pipeline~\citep{Zhang20}. 
Strong radio frequency interference (RFI), both narrow-band and short-duration broadband types, was identified and masked using the \emph{\sc presto} routine \emph{\sc rfifind}.
The data were then dedispersed over a DM range of $0-1000$ pc\,cm$^{-3}$, with the steps determined by \emph{\sc ddplan.py}. 
The DM upper limit of 1000 pc\,cm$^{-3}$ was chosen to conservatively exceed the maximum estimated DM among the four clusters$-$approximately 194.9 pc\,cm$^{-3}$ for Trumpler 20$-$while providing a safety margin to account for uncertainties in Galactic electron density models and open cluster distance estimates.
Single-pulse candidates with sigma~\footnote{The definition of ``sigma'' used in \emph{\sc presto} is: $\Sigma ({\rm signal} - {\rm background\_level})/({\rm RMS} \centerdot \sqrt{\rm baxcar\_width})$. This represents the sum of the signal above the background level, divided by the product of the noise root-mean-square (RMS) and the square root of the boxcar filtering parameters. An advantage of this definition is that it yields approximately the same sigma value for a given pulse, regardless of the level of downsampling applied to the input time series, as long as the pulse remains resolved. It is important to note that this definition differs from the conventional signal-to-noise ratio (S/N) calculated using the radiometer equation~\citep{Lorimer04handbook}, which typically refers to the peak signal divided by the noise RMS.} exceeding 7
were identified using the \emph{\sc single\_pulse\_search.py} routine
and visually inspected.  
To minimise false positives 
caused by structured and strong RFI, which indeed dominate the number of identified candidates in our search, a candidate was considered a real detection only if it exhibited a plausible sweep across the dedispersed frequency-time plane. For multibeam data, the signal had to be detected in no more than three adjacent beams~\footnote{The three-beam threshold accounts for the Parkes multibeam receiver's feed configuration, where sources near beam intersections may appear in adjacent beam sidelobes, as demonstrated by ``Lorimer burst'' initial three-beam detection~\citep{Lorimer07}. In contrast, terrestrial RFI like perytons~\citep{Burke-Spolaor11} typically exhibit detection across almost all 13 beams.}.
%

Periodicity searches were also performed using \emph{\sc presto},  following a similar preprocessing pipeline.  
RFI was again rejected and marked using \emph{\sc rfifind}, and trial DMs up to $1000\,\,$pc\,cm$^{-3}$ were determined with \emph{\sc ddplan.py}. 
The concatenated data were then dedispersed at each trial DM using \emph{\sc prepsubband}, which also removes Doppler shifts induced by the Earth's motion.   
The resulting time series were Fourier transformed using \emph{\sc realfft} package,
and periodic signals were identified in the Fourier domain suing \emph{\sc accelsearch} with \emph{\sc -zmax}=200 to search for the potentially accelerated systems.
Candidates with significance less than 4$\sigma$ (from incoherent power summation) and with maximum sigma occurring at a DM of 0\,cm$^{-3}$pc were excluded. Multiple detections, such as harmonics or the same barycentric period at multiple DMs, were combined using \emph{\sc accel\_sift.py}. 
Finally, candidates with sigma greater than 8 suggested by \emph{\sc accel\_sift.py} were folded using \emph{\sc prepfold} for further visual inspection.

\section{Results and Discussion} \label{sec:results}

\begin{table*}
\scriptsize
    \centering
\begin{tabular}{ccc|cccccllcc}
\toprule
Cluster   &  DM$_{\rm YMW16}$ &  DM$_{\rm NE2001}$  &  Source          &  Source &  SPS  & PS & DM$^a$   &  R.A.$^a$     &  Decl.$^a$ & D$_{\rm cc}$$^b$  & Age \\ 
Name     &  (pc\,cm$^{-3}$) & (pc\,cm$^{-3}$)  & Name    & type    &  sigma &  sigma &  (pc\,cm$^{-3}$) &   (deg)   &  (deg) &  (deg) & (yr) \\
    \hline   
Theia 1661  &  82.2  & 113.0  & PSR J1749$-$2629  & known & 8.0 & 38.5 & 409 & 267.297       & -26.486     & 1.42 & 1.23$\times 10^7$         \\ 
            &        &        & PSR J1750$-$2536  & known & - & -  & 179 & 267.639      & -25.612     & 0.47 & -    \\ 
            &        &        & RRAT J1749$-$25   & new   &  7.9/7.5 & -  & 56 $\pm$ 10  & 267.325$^*$   &  -24.994$^*$  & 0.76 & -    \\ 
\hline
NGC 6259  &   76.6  &  71.6   &  PSR J1659$-$4439  & known & - & 26.3 & 536 &  254.914  & -44.650 & 0.20 & 2.04$\times 10^8$  \\ 
          &         &         &  PSR J1700$-$4422   & known & - & 9.0 & 405 & 255.224      & -44.374  & 0.31 & 2.99$\times 10^8$\\ 
          &         &         &  PSR J1702$-$4428   & known & -  & 16.8  &  395 & 255.719     & -44.468   & 0.43 & 1.02$\times 10^7$\\ 
          &         &         &  RRAT J1702$-$44   & new & 8.0 & - & 281 $\pm$ 30 & 255.681 & -44.722  & 0.35 & -  \\ 
\hline
Trumpler 20 &  194.9  & 166.8  &  RRAT J1237$-$60   & new & 7.9 & - & 125 $\pm$ 20  & 189.288  & -60.818  & 0.35  & - \\     
\hline
\end{tabular}
    \caption{Characteristics of detected pulsars and RRAT candidates in the direction of the observed open clusters. 
    $^a$ For known pulsars, DMs and positions are from the ATNF pulsar catalogue~\citep{Manchester05}. For newly discovered RRAT candidates, DM values and positions were estimated from the detection data. 
    $^b$ D$_{\rm cc}$ refers to the distance from the source to the cluster centre.
    $^*$ is the sigma-weighted refined position of RRAT J1749$-$25 derived from two separate detections targeting at RA = 267.590$^{\circ}$, Dec = $-$25.077$^{\circ}$ (J2000) and  RA = 267.046$^{\circ}$, Dec = $-$24.906$^{\circ}$ (J2000).}
    \label{tab:cand_list}
\end{table*}

\begin{figure*}[!htb]
\centering
 \begin{tabular}{cc}
SPC2: DM $\sim$ 56 pc\,cm$^{-3}$ &
SPC3: DM $\sim$ 56 pc\,cm$^{-3}$  \\
\includegraphics[width=0.32\linewidth]{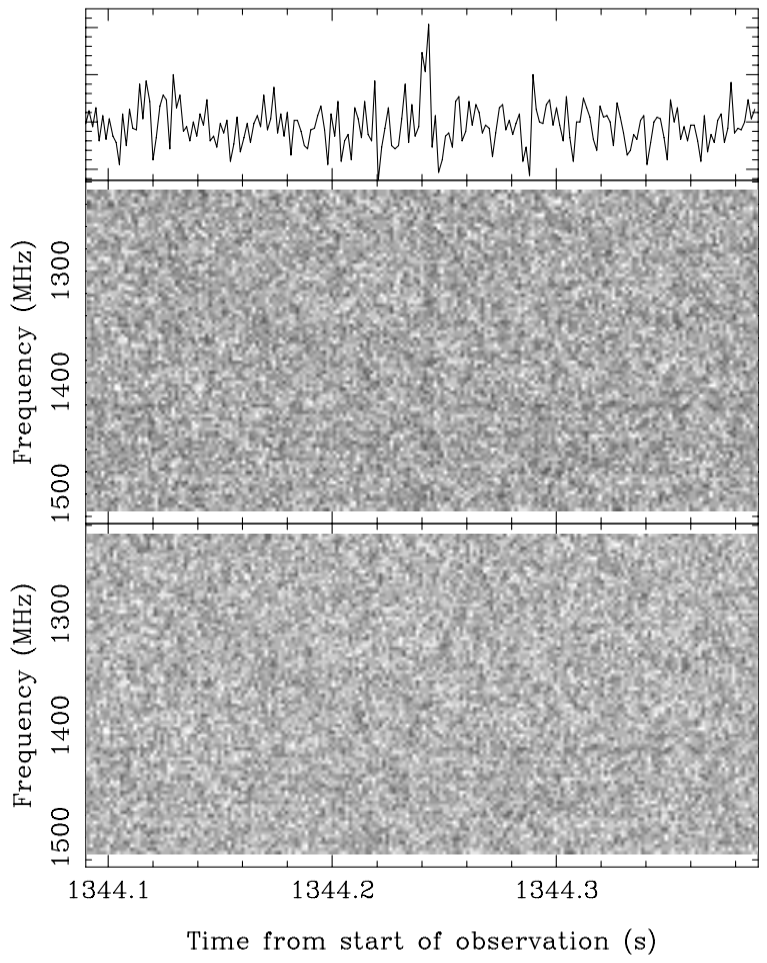} & \includegraphics[width=0.32\linewidth]{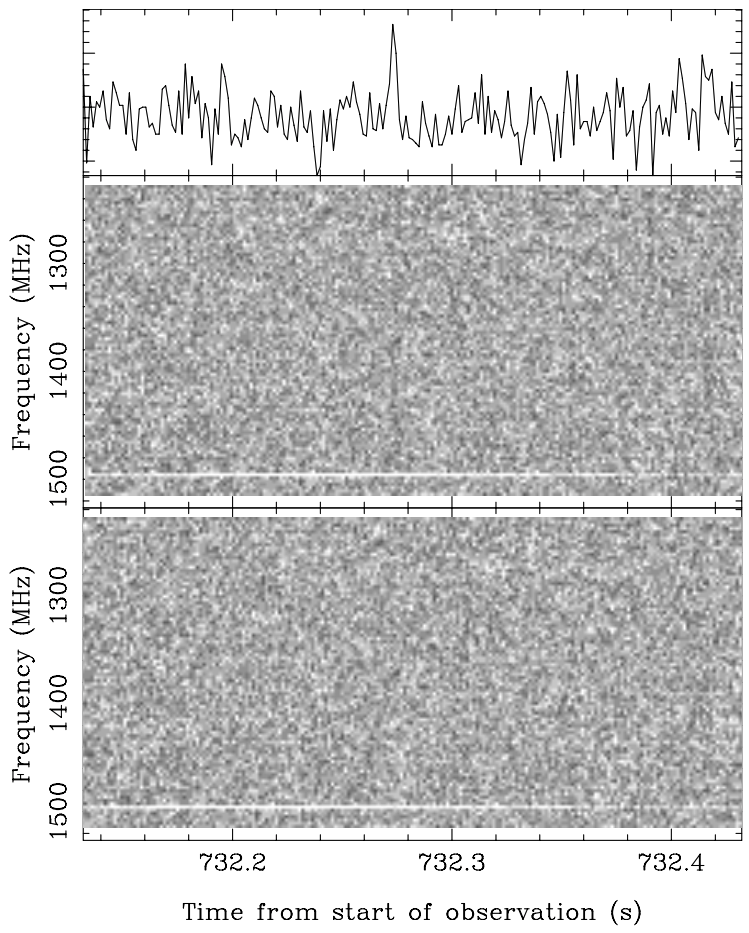}  \\
\end{tabular}
\begin{tabular}{ccc}
SPC1: DM $\sim$ 414 pc\,cm$^{-3}$ &
SPC4: DM $\sim$ 281 pc\,cm$^{-3}$ &
SPC5: DM $\sim$ 125 pc\,cm$^{-3}$  \\
\includegraphics[width=0.32\linewidth]{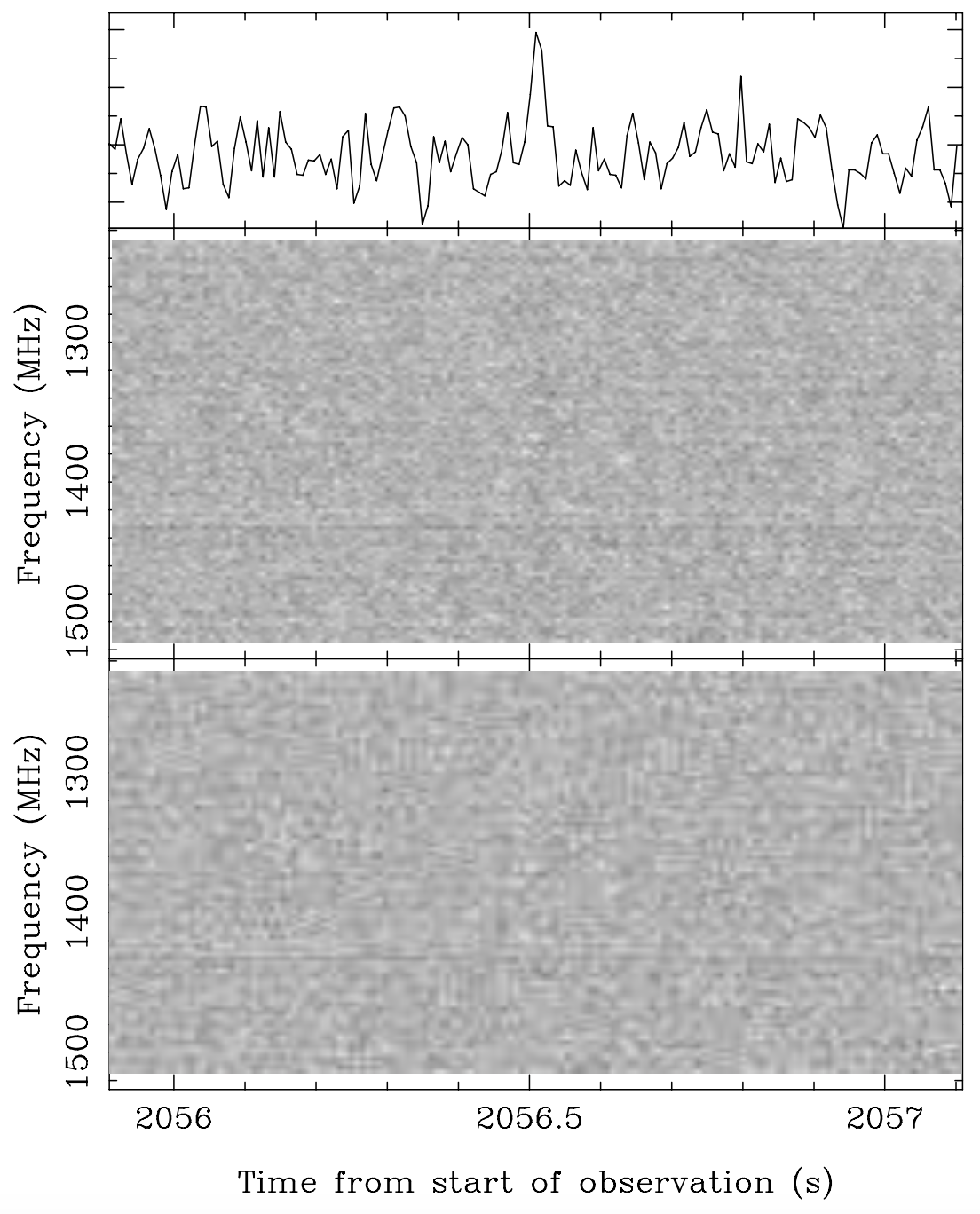} &
\includegraphics[width=0.32\linewidth]{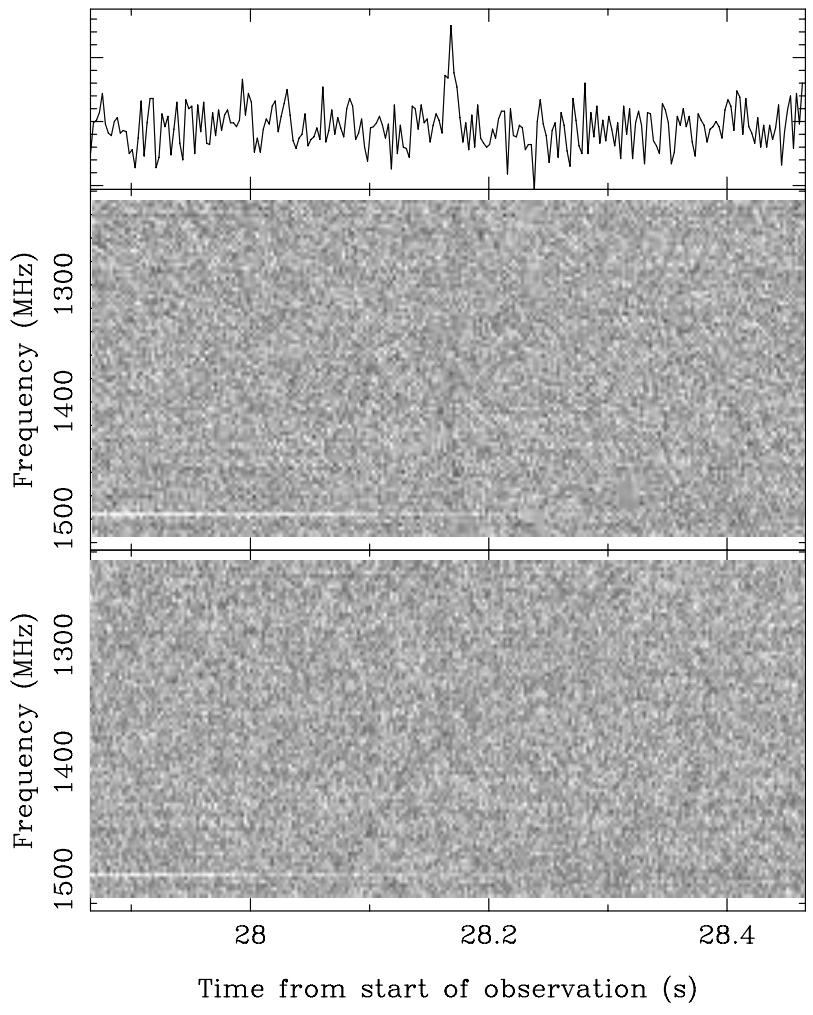} &
\includegraphics[width=0.32\linewidth]{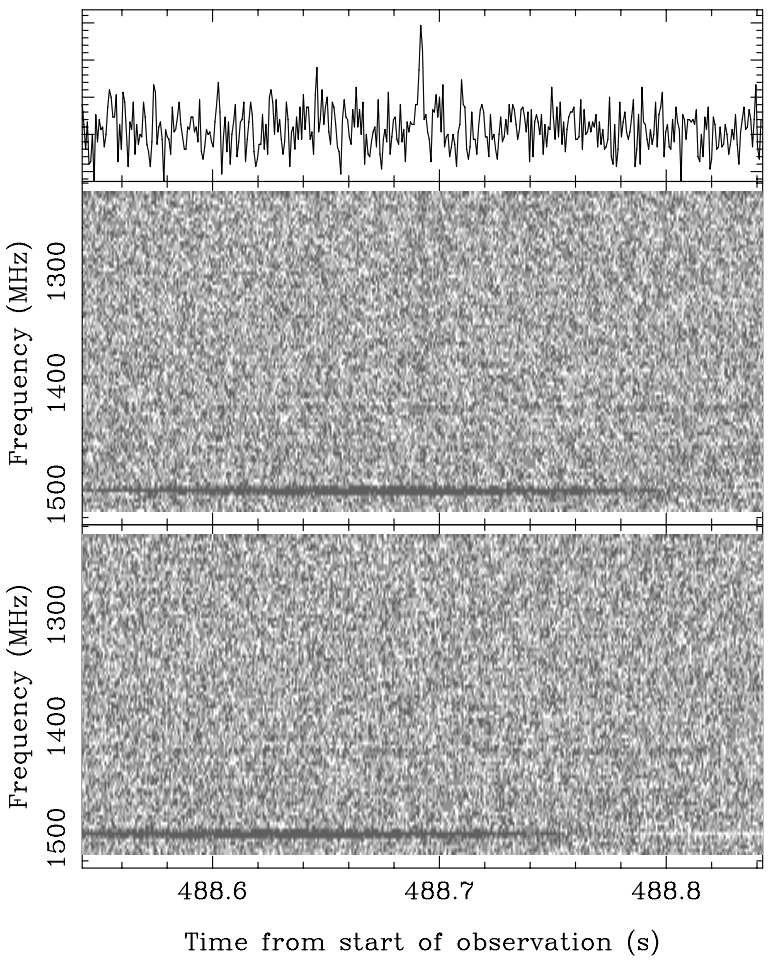}\\
 \end{tabular}
\caption{The profiles and dynamic frequency-time spectra of the detected single pulses. {\bf Top panels:} Two single pulses (SPC2 and SPC3) from RRAT J1749$-$25 with sigma values of 7.9 and 7.5 detected in the direction of Theia 1661. {\bf Bottom panels (left to right)}: Single pulse (SPC1) from known PSR J1749$-$2629 (sigma $\sim$ 8.0) in the direction of Theia 1661; single pulse (SPC4) from RRAT J1702$-$44 (sigma $\sim$ 8.0) in the direction of NGC 6259; and single pulse (SPC5) from RRAT J1237$-$60 (sigma $\sim$ 7.9) in the direction of Trumpler 20.
For each subfigure, the bottom panel shows the frequency–time plane, the central panel displays the event after being dedispersed at the optimized DMs, while the top panel is the integrated pulse profile using an arbitrary flux scale. The time axis indicates the time since the start of the observation, with a resolution of 7.5\,ms,  1.5\,ms, 1.0\,ms, 3.0\,ms and 0.75\,ms for SPC1, SPC2, SPC3, SPC4 and SPC5, respectively. The frequency resolution is 3\,MHz.
}
\label{fig:spc}
\end{figure*}

Through the single-pulse search pipeline, five plausible candidates were identified from a total of 75.02 hours of archival Parkes observations. Their pulse profiles and dynamic spectra are shown in Figure~\ref{fig:spc}, and the measured properties are listed in Table~\ref{tab:cand_list}. 

In the observations of 47.74 hours toward Theia 1661, a single pulse (hereafter SPC1) with an sigma of 8.0 was detected at a DM of $\sim 414$ pc\,cm$^{-3}$ in the observation pointing centered at RA = 267.331$^{\circ}$, Dec = $-$26.568$^{\circ}$ (J2000) on MJD 50593.
Two single pulses (hereafter SPC2 and SPC3) with sigmas of 7.9 and 7.5 were detected at a DM of $\sim 56$ pc\,cm$^{-3}$ in the observations targeting at RA = 267.590$^{\circ}$, Dec = $-$25.077$^{\circ}$ (J2000) on MJD 50593, and RA = 267.046$^{\circ}$, Dec = $-$24.906$^{\circ}$ (J2000) on MJD 51151, respectively. 
These two observations~\footnote{Observations used beams 6 and 7 of the Parkes 21-cm Multibeam receiver~\citep{Staveley-Smith96MB}, which have half-power beamwidths of $\sim 14.1$ arcmin at 1.4\,GHz.} have a half-power beamwidth (HPBW) of $\sim 14.1$ arcmin. The separation between these two pointings is $\sim 31.3$ arcmin, slightly exceeding twice the HPBW, suggesting that SPC2 and SPC3 could originate from the same source.
Based on their relative sigmas, we estimate a refined position of RA = 267.325$^{\circ}$, Dec = $-$24.994$^{\circ}$ (J2000) for the potential source.

In the direction of NGC 6259 of 10.05 hours observations, a single pulse (hereafter SPC4) with an sigma of 8.0 was detected at a DM of $\sim 281$ pc\,cm$^{-3}$ in an observation centered at RA = 255.681$^{\circ}$, Dec = $-$44.722$^{\circ}$ (J2000) on MJD 50593. 
Similarly, toward Trumpler 20 of 11.67 hours observations, one single pulse (hereafter SPC5) with an sigma of 7.9 was detected at a DM of $\sim 125$ pc\,cm$^{-3}$ in a pointing centered at RA = 189.288$^{\circ}$, Dec = $-$60.818$^{\circ}$ (J2000), also on MJD 50593.
For all five pulses, we examined data from other beams of the multibeam observation, but found no additional detections near their respective arrival times.

\begin{figure*}[!htb]
\centering
\begin{tabular}{cc}
PSR J1749$-$2629 & PSR J1659$-$4439 \\
\includegraphics[width=0.45\linewidth]{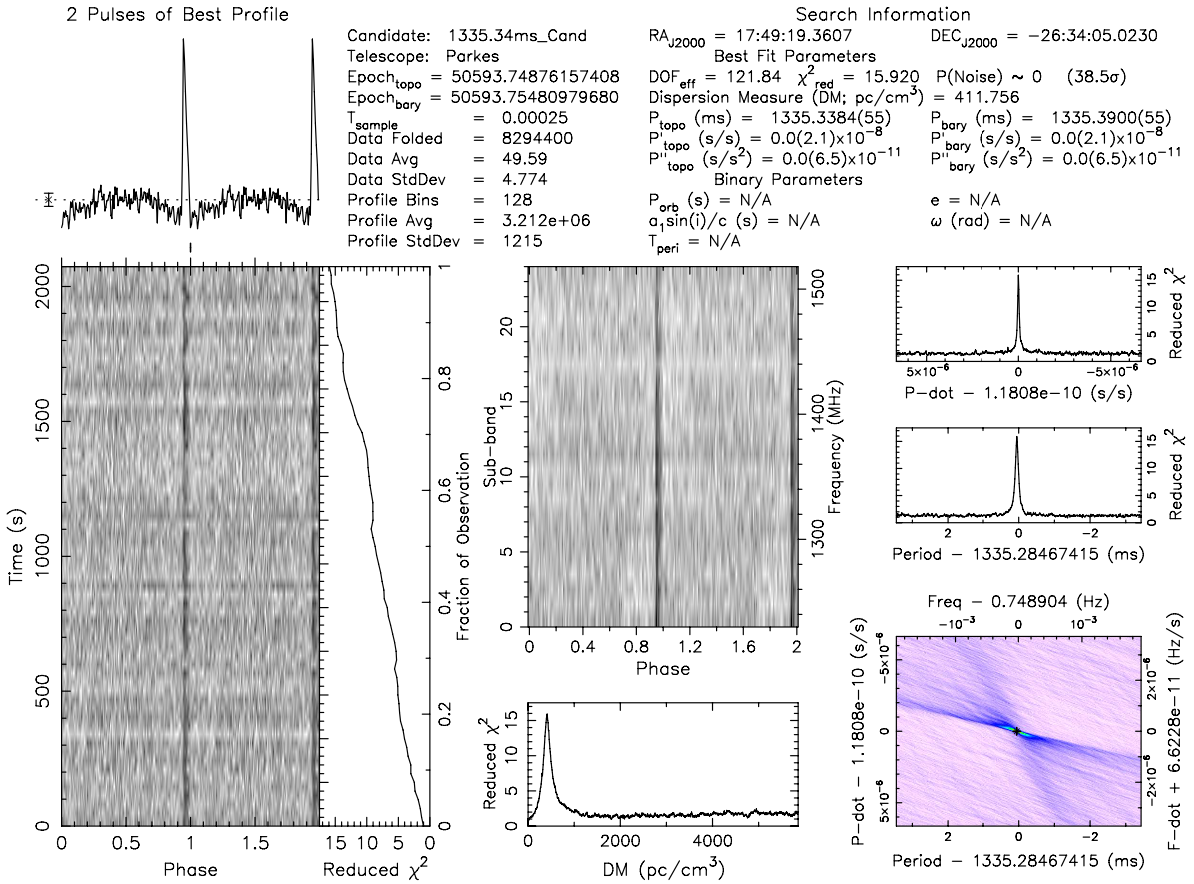} & \includegraphics[width=0.45\linewidth]{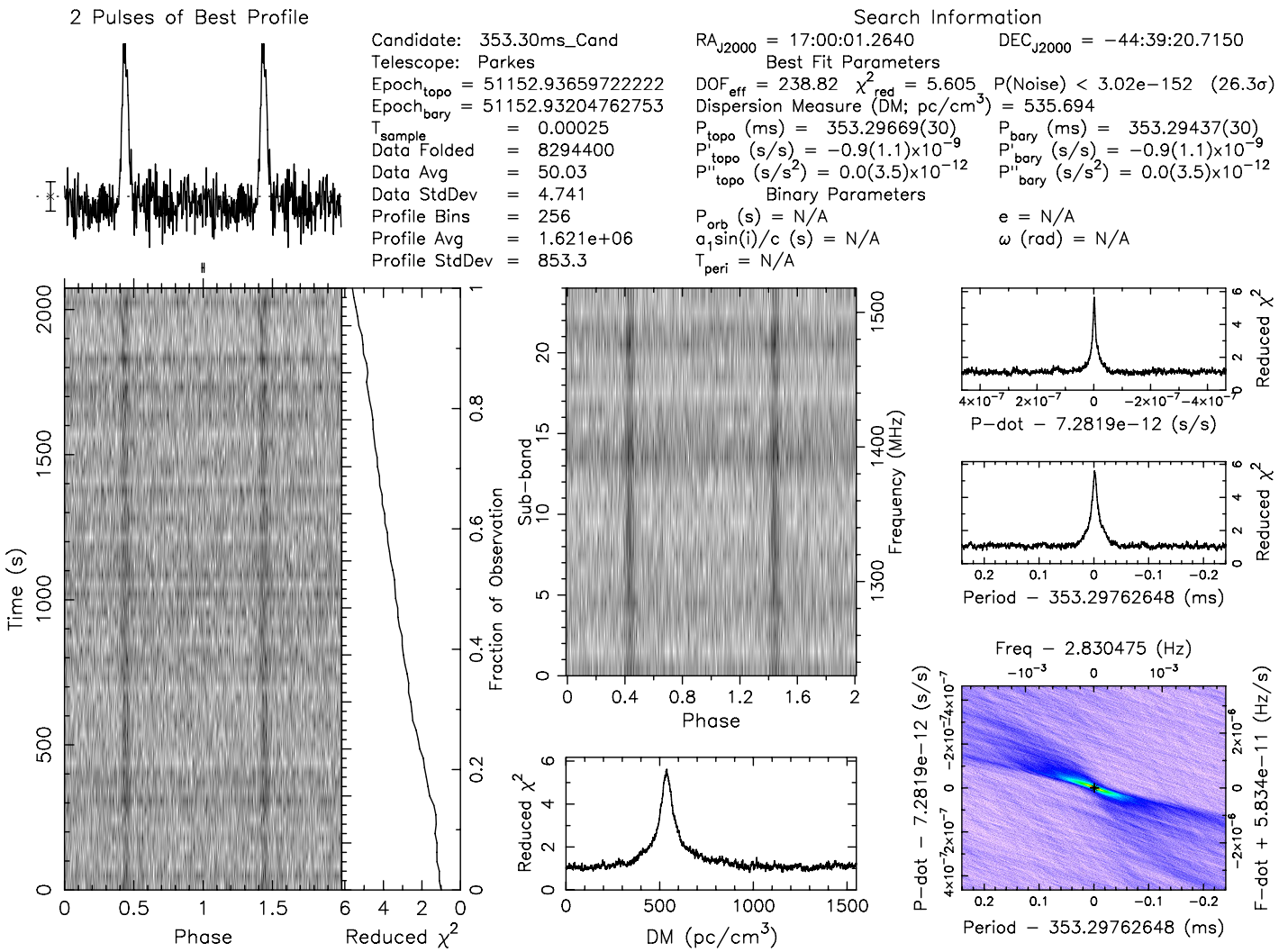} \\
\end{tabular}
\begin{tabular}{cc}
 PSR J1700$-$4422 &  PSR J1702$-$4428 \\
\includegraphics[width=0.45\linewidth]{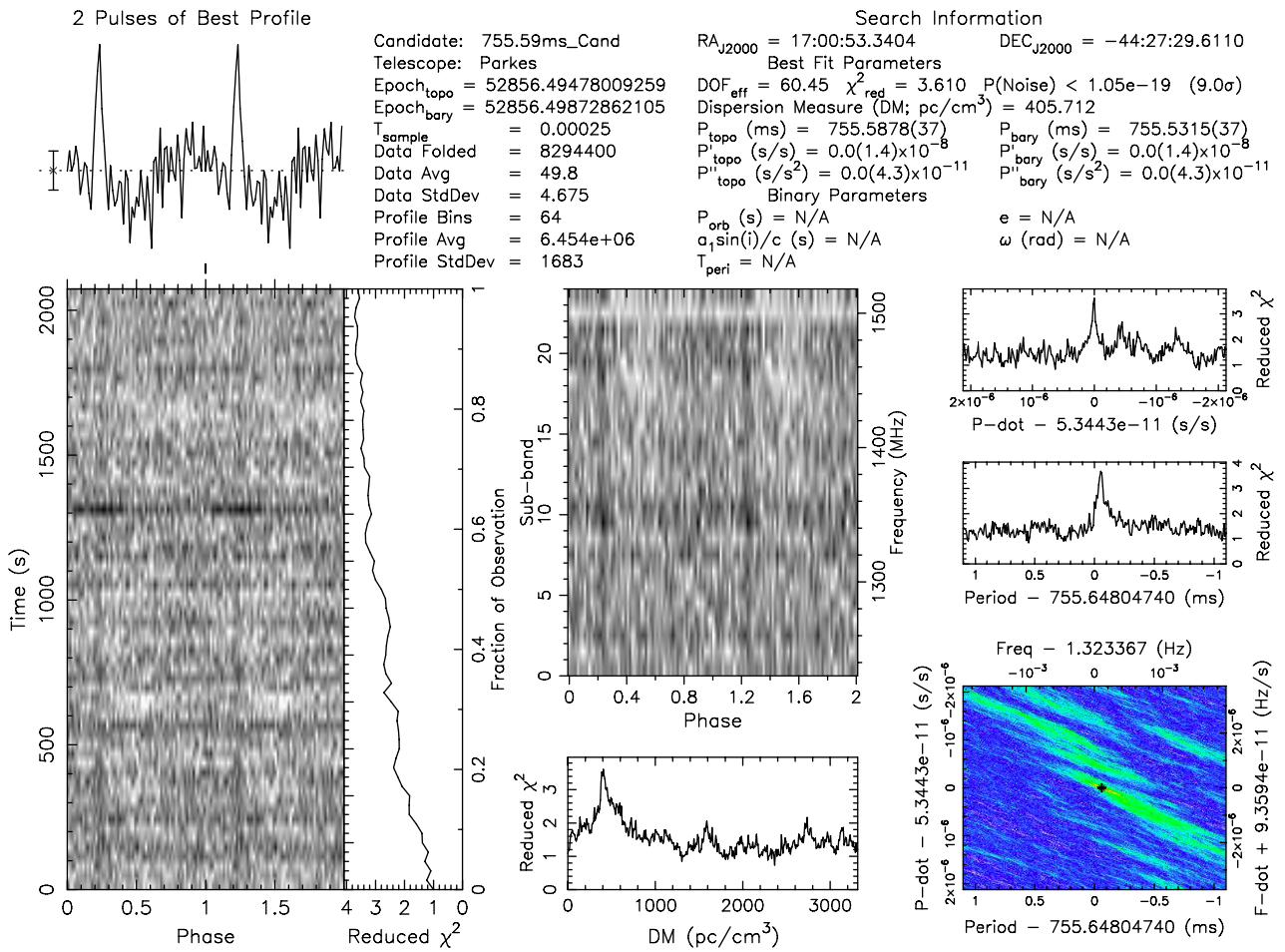} & \includegraphics[width=0.45\linewidth]{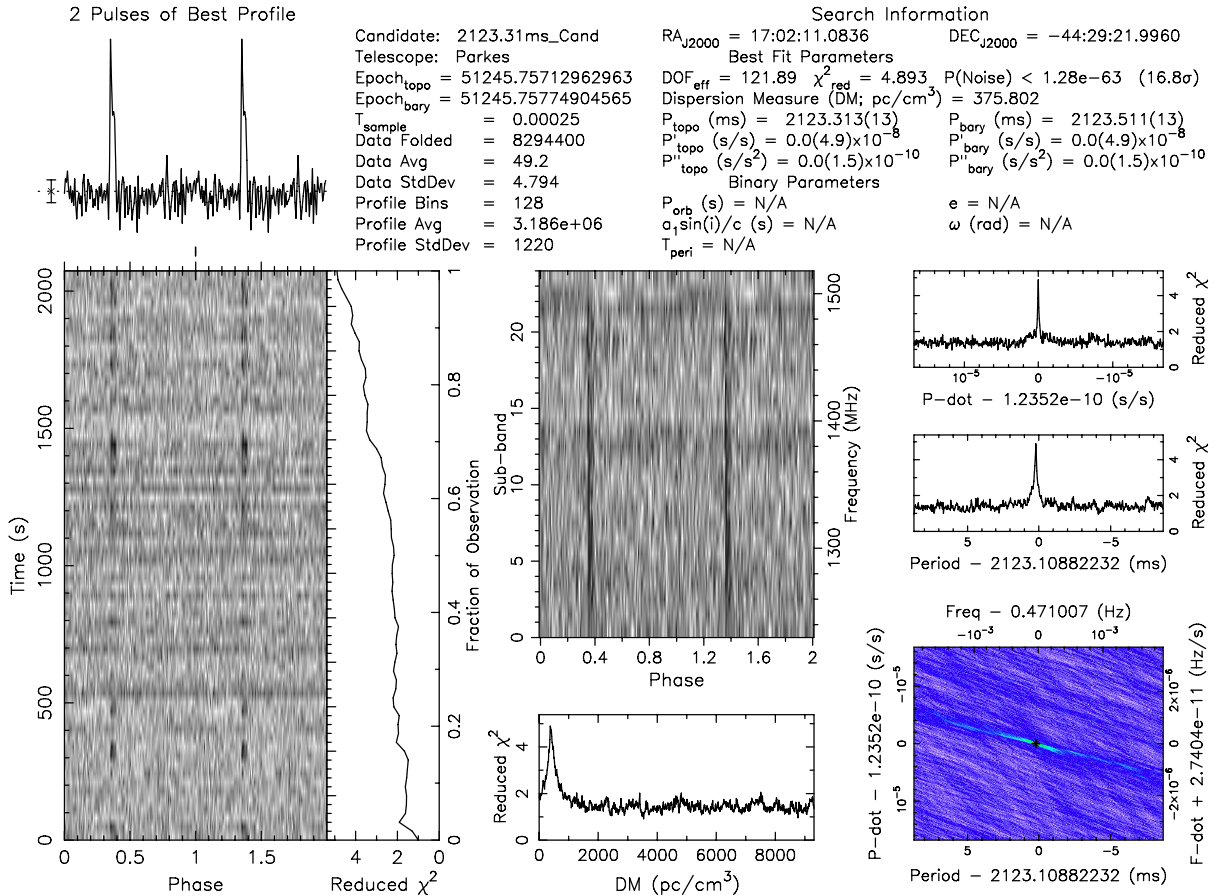} \\
\end{tabular}
\begin{tabular}{c}
 PSR J1750$-$2536 \\
\includegraphics[width=0.45\linewidth]{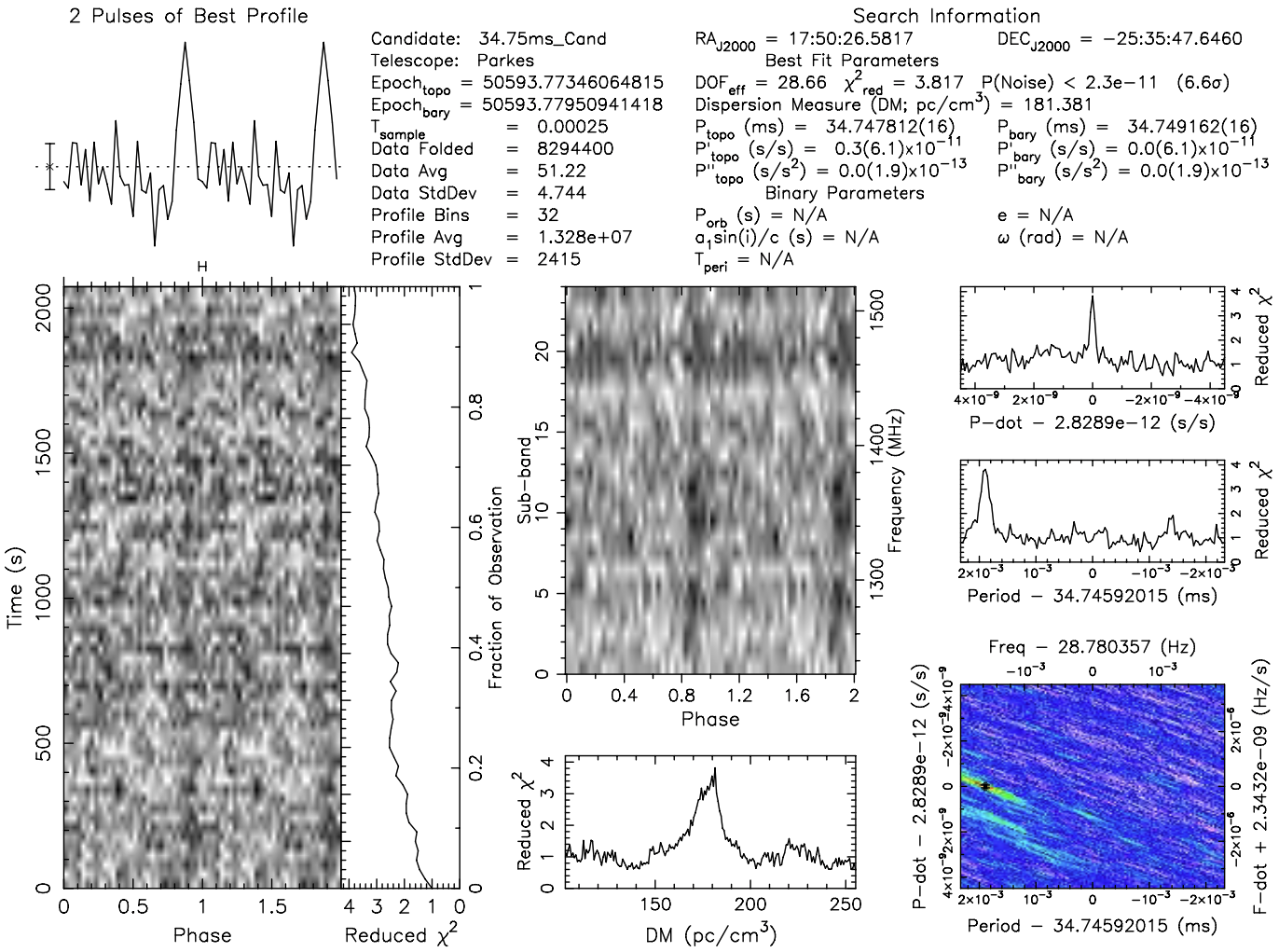} \\
 \end{tabular}
\caption{Examples of \emph{\sc presto} search diagnostic plots of the archival data sets, showing the periodicity detections of PSRs J1749$-$2629 (which also contains the single-pulse detection SPC1), J1659$-$4439, J1700$-$4422  and J1702$-$4428. The periodicity search result for PSR J1750$-$2536 (sigma $\sim 6.6$) did not meet our detection threshold (sigma $> 8$), but it was identified through cross-matching with known pulsars in the ATNF catalogue~\citep{Manchester05}.}
\label{fig:psr_ps}
\end{figure*}

In addition, our periodicity search detected four pulsars with sigmas above the detection threshold of eight in multiple search-mode data sets, examples of each pulsar detection are shown in Figure~\ref{fig:psr_ps}. 
By comparing their measured DMs, periods, and pointing directions with the Australia Telescope National Facility (ATNF) pulsar catalogue\footnote{ATNF Pulsar Catalogue v2.6.0: \url{https://www.atnf.csiro.au/research/pulsar/psrcat}}~\citep{Manchester05}, we identified them as the known pulsars PSRs J1749$-$2629, J1659$-$4439, J1700$-$4422 and J1702$-$4428. 
Notably, the observation in which SPC1 was detected also includes a periodic detection of PSR J1749$-$2629~\footnote{Note that its pointing of RA = 267.331$^{\circ}$, Dec = $-$26.568$^{\circ}$ (J2000) differs slightly from the refined timing position of PSR J1749$-$2629, which is RA = 267.297, Dec = $-$26.486 (J2000)~\citep{Morris02}}(Figure~\ref{fig:psr_ps}). The estimated DM of SPC1 ($414 \pm 30$ pc\,cm$^{-3}$) is consistent with the catalogued DM of PSR J1749$-$2629 ($\sim409$ pc\,cm$^{-3}$), strongly suggesting that SPC1 is a single pulse from this known source. 
No significant periodic signals were found in the data sets where SPC2$–$SPC5 were detected.

We also cross-matched the spatial locations of Theia 1661, NGC 6259, Pismis 3, and Trumpler 20 with known pulsars in the ATNF catalogue~\citep{Manchester05}. One additional pulsar, PSR J1750$-$2536, lies in the direction of Theia 1661 but was not detected in our reprocessed search-mode data sets. The discovery of PSR J1750$-$2536 required novel methods involving distributed computing via EINSTEIN@HOME~\citep{Knispel13}, as standard \emph{\sc presto} searches yielded an sigma of only $\sim 6.6$ in our data. Consequently, its periodicity search result did not meet our detection threshold (sigma $> 8$), though we still present it in Figure~\ref{fig:psr_ps}. 
PSR J1750$-$2536 is not associated with any of the newly detected single pulses.

As summarized in Figure~\ref{fig:pointing} and Table~\ref{tab:cand_list}, in the direction of Theia 1661, NGC 6259, or Trumpler 20, our analysis identifies five known pulsars (PSR J1749$-$2629, J1750$-$2536, J1659$-$4439, J1700$-$4422, and J1702$-$4428) and three new RRAT candidates: J1749$-$25, which emitted two single pulses (SPC2 and SPC3); J1702$-$44 (SPC5) and J1237$-$60 (SPC4), which emitted a single pulse each. 
   
However, the DMs of PSR J1749$-$2629, J1659$-$4439, J1700$-$4422, J1702$-$4428, and RRAT J1702$-$44, are significantly larger than the expected DMs of the related clusters. 
This strongly suggests that they are background pulsars rather than cluster members.
In contrast, the estimated DM of RRAT candidate J1749$-$25 ($56\pm 10$ pc\,cm$^{-3}$) is comparable to the predicted DMs for Theia 1661: 82.2 and 113.0 pc\,cm$^{-3}$ based on the YMW16 and NE2001 electron density models, respectively.
Likewise, the estimated DM of RRAT J1237$-$60 ($125\pm 20$ pc\,cm$^{-3}$) is reasonably close to the predicted DMs of Trumpler 20: 194.9 (YMW16) and 166.8 (NE2001) pc\,cm$^{-3}$.
These consistencies suggest that both RRATs may originate from within or near their related clusters.

The DM of PSR J1750$-$2536 ($179$ pc\,cm$^{-3}$) is approximately twice the expected DM of Theia 1661, making it more likely a background source.
Nevertheless, given the uncertainties in the distance estimation of Theia 1661~\citep{Hunt24} and the inherent limitations of current electron density models, we cannot definitively rule out the possibility that PSR J1750$-$2536 is a member of Theia 1661.

\section{Conclusions} \label{sec:con}

We reprocessed 224 archival Parkes observations (totalling 75.02 hours) targeting four old open clusters proposed to be similar to NGC 6791$-$Theia 1661, NGC 6259, Pismis 3, and Trumpler 20. 
By cross-matching our results with known pulsars in the ATNF catalogue~\citep{Manchester05}, we identified five known pulsars and three new RRAT candidates in the direction of these clusters.
Comparing the measured DMs of these sources with the clusters DMs estimated using the YMW16~\citep{Yao17} and NE2001~\citep{ne2001} electron density models, we conclude that PSR J1749$-$2629, J1659$-$4439, J1700$-$4422, J1702$-$4428, and RRATs J1702$-$44 are likely background sources. 
The association between PSR J1750$-$2536 and cluster Theia 1661 remains ambiguous.
However, RRATs J1749$-$25 and J1237$-$60 are strong candidate members of Theia 1661 and Trumpler 20, respectively.

RRATs represent a subclass of pulsar characterised by their sporadic radio emission, making them particularly challenging to detect through conventional periodicity searches~\citep{McLaughlin06, Bagchi13}. 
RRATs J1749$-$25 and J1237$-$60 were identified through only two and one single pulse(s), respectively, each with an sigma of $\sim 8$, distinguishing them from typical pulsars. 
However, recent studies suggest that some Galactic radio sources may emit only one-off pulses~\citep{Keane16,Zhang24_Parkes}, and our detection of SPC1 with similar sigma was later confirmed as a pulse from PSR J1749$-$2629, which supports the astrophysical nature of these faint single-pulse detections.
Due to their relatively low escape velocities, open clusters may have difficulty retaining typical pulsars that receive large natal kicks.  Nevertheless, they might provide valuable environments for discovering unusual neutron star populations~\citep{Liu25}.
The detection of RRATs possibly associated with old open clusters may offer important insights into stellar evolution and neutron star retention in such environments.
Additionally, PSR J1750$-$2536 has been proposed as a rare intermediate-mass binary pulsar~\citep{Knispel13}. If confirmed as a member of Theia 1661, it would provide further constraints on binary evolution and neutron star retention in open clusters.

RRATs J1749$-$25 and J1237$-$60 exhibit weak and extremely sporadic emission in Parkes data. 
PSR J1750$-$2536 was originally discovered through novel search techniques and is not easily detected through standard methods~\citep{Knispel13}. Even over a decade later, no refined timing solution has been obtained~\citep{Spiewak22}.  
We have proposed more monitoring of these sources with the Parkes UWL receiver.
However, follow-up observations with more sensitive radio telescopes, such as MeerKAT or the Square Kilometre Array (SKA), will be critical for investigating the potential association of these sources with their host clusters. 
A dedicated high-sensitivity monitoring campaign would have a better chance to reveal additional pulses, better-integrated profiles, and with a long baseline, enabling precise timing solutions to determine positions, distances, and proper motions—parameters essential for definitively confirming or rejecting cluster membership.
Finally, deeper searches of these old and massive open clusters may uncover additional pulsar candidates, potentially establishing open cluster pulsars as a distinct and systematically observable population.

\section*{Acknowledgments}
This research has been partially funded by the International Partnership Program of Chinese Academy of Sciences for Grand Challenges (114332KYSB20210018), the National SKA Program of China (2022SKA0130100), the National Natural Science Foundation of China (grant Nos. 12041306,12273113,12233002,12003028,12321003,12422307, 12373053), the CAS Project for Young Scientists in Basic Research (Grant No. YSBR-063), and the ACAMAR Postdoctoral Fellow. \\


\bibliography{sample631}{}
\bibliographystyle{aasjournal}



\end{document}